\documentclass[10pt]{article}
\newcommand{\be}{\begin{equation}}
\newcommand{\ee}{\end{equation}}
\newcommand{\bea}{\begin{eqnarray}}
\newcommand{\eea}{\end{eqnarray}}

\def\id{\protect{{1 \kern-.28em {\rm l}}}}

\def\qnc{{Q_{N_c}^\alpha}}
\def\tqnc{{\tilde Q}_{N_c}^\alpha}
\def\qnc1{Q_{N_c+1}^\alpha}
\def\tqnc1{{\tilde Q}_{N_c+1}^\alpha}

\def\appendix#1{
  \addtocounter{section}{1}
  \setcounter{equation}{0}
  \renewcommand{\thesection}{\Alph{section}}
  \section*{Appendix \thesection\protect\indent \parbox[t]{11.715cm} {#1}
}
  \addcontentsline{toc}{section}{Appendix \thesection\ \ \ #1}
  }

\begin{document}

\begin{titlepage}

\hfill\hbox to 3cm {\parbox{4cm}{
NSF-KITP-04-120 \\
}\hss}

\vspace{.5cm}

\begin{center}

\mbox{\large\bf Geometric Transitions,}

\vspace{.15cm}

\mbox{\large\bf  Non-Kahler Geometries and String
Vacua}

\vspace{1cm}

{Katrin Becker$^a$,~ Melanie Becker$^b$,~ Keshav Dasgupta$^c$,~ Radu
Tatar$^d$}

\vspace{1cm}

{\it ${}^a$ Department of Physics, University of Utah\\}
{\it Salt Lake City, UT 84112, USA}

\vspace{.5cm}

{\it ${}^b$ Department of Physics, University of Maryland \\}
{\it College Park, MD 20742, USA}

\vspace{.5cm}

{\it ${}^c$ Department of Physics, Stanford University \\}
{\it Stanford, CA 94305, USA}

\vspace{.5cm}

{\it ${}^d$ Theoretical Physics Group, LBL Berkeley \\}
{\it Berkeley, CA 94720, USA}

\vspace{.5cm}

{\it ${}^d$Kavli Institute for Theoretical Physics \\}
{\it University of California \\}
{Santa Barbara , CA 93106-4030, USA}

\end{center}

\vspace{1.5cm}

\begin{abstract}
\noindent We summarize an explicit construction of a duality cycle for
geometric transitions in type II and heterotic theories. We
emphasize that the manifolds with torsion constructed with this
duality cycle are crucial for understanding different phenomena
appearing in effective field theories.
\end{abstract}

\end{titlepage}

\section{Introduction}

The connections between string theory and realistic supersymmetric
gauge theories have been extensively studied in the last years.
One of the approaches is the one taken by Vafa \cite{vafa1} that
is based on the duality between open topological strings on a
Lagrangian submanifold and closed topological strings on a
resolved conifold. This has been extended to the type IIB theory
in \cite{civ,eot} where the open string side is described by D5
branes wrapping a two cycle of a resolved conifold and the closed
string side is a warped deformed conifold with fluxes. The open
string side captures the far IR behavior of the gauge theory. The
full picture which studies the UV as well as IR behavior was
described in reference \cite{katz} where the cascading from the UV
to the IR is shown to arrive from an infinite sequence of flop
transitions.

The first goal of this note is to describe the type IIA
transition in detail, by considering D6 branes wrapped on three
cycles inside a non-K\"ahler deformation of the deformed conifold
\cite{trans1}. This corresponds to the open string picture. The
closed string dual is a compactification with RR  fluxes on
another non-K\"ahler manifold with $dJ \ne 0$ and $d \Omega \ne
0$ and with a superpotential
\begin{equation}
W_{IIA} = \int (J+i B) \wedge d \Omega.
\end{equation}

The second goal is to extend the cycle of geometric transitions
to type I and heterotic strings. The dual corresponds to either a
type I string compactified on a non-K\"ahler but complex manifold
(observe that the type IIA manifolds were non-complex) with the
superpotential \cite{prok,ccdl}
\begin{equation}
W_{I} = \int (H_{RR} + i d J) \wedge \Omega,
\end{equation}
or to the heterotic string compactified on a non-K\"ahler but
again complex manifold with the superpotential \cite{prok,ccdl}
\begin{equation}
W_{het}=\int(H+i d J)\wedge \Omega.
\end{equation}
Here $H$ is the usual three form of the heterotic theory that
satisfies $dH = {\rm tr} ~R \wedge R - \frac{1}{30} \mbox{tr}~F
\wedge F$ and $H_{RR}$ is the RR three form that is S-dual to $H$.

One important result of our work is that we identify a geometric
transition for both the type I and heterotic string theories. We
will be able to specify the backgrounds in the type I and
heterotic theories on both sides of the duality. This suggests
the possibility of having a gravity description for wrapped D5
branes (for type I) or wrapped NS5 branes (for heterotic). Our
work also provides an alternative picture to the Landscape
distribution of string vacua \cite{kklt,douglas}.

\section{Type IIA Superstrings and Non-Calabi-Yau Manifolds}

Geometric transitions are examples of generalized AdS/CFT
correspondences which relate D-branes in the open string picture
and fluxes in the closed string picture. There are several types
of geometric transitions depending on the framework in which we
formulate them. The type IIB geometric transition starts with
$D5$ branes wrapping a $P^1$ of a resolved conifold. The
corresponding metric is given in \cite{pando} as
\begin{equation}
\label{2b} ds^2 = (dz + \Delta_1~{\rm cot}~\theta_1 ~dx +
\Delta_2~{\rm cot}~\theta_2 ~dy)^2 + \vert dz_1\vert^2 + \vert
dz_2\vert^2,
\end{equation}
where we have
replaced the metric of two spheres of the resolved conifold by two
tori with complex structures $\tau_1$ and $\tau_2$. The complex one forms
$dz_i,~ i = 1, 2$
are therefore defined as
\begin{equation}
dz_1 = dx - \tau_1~d\theta_1, ~~~~~~ dz_2 = dy - \tau_2~d\theta_2.
\end{equation}

We now want to apply the result of \cite{syz} which tells us that
for a manifold admitting a $T^3$ structure, and in the limit of a
large complex structure the mirror manifold is obtained by
performing three T-dualities on the $T^3$ torus \footnote{This is
a different type of T-duality than the one considered in
\cite{dot,rr,llt} where the result of a single T-duality was a
brane configuration in type IIA.}. In our case, the large complex
structure is obtained by boosting \footnote{This can also be
viewed as if we had introduced new complex structures on the two
tori. There is a subtlety related to whether these complex
structures are integrable or not. Considering only the integrable
complex structures leads us very close to the right mirror metric,
which can nevertheless be obtained by choosing non-integrable
complex structures. This discussion has appeared in
\cite{trans1}.}
\begin{equation}
dz_i \rightarrow dz_i + f_i~d\theta_i,~~~f_i \rightarrow \infty,~~i=1,2.
\end{equation}
In the presence of a NS field with components $b_{x\theta_1}$ and
$b_{y\theta_2}$, we have explicitly performed the mirror
tranformation in \cite{trans1}. The outcome after the mirror
transformation
 is the following metric \footnote{There is a subtlety that we should mention here.
 The mirror rules of \cite{syz} tell
us that we should take a limit of large complex structures. On the other hand
 geometric transitions occur in exactly the opposite limit. Therefore naively applying
\cite{syz} we do not get the right answer. The correct answer was
derived in \cite{trans1} by performing a set of coordinate
transformations.}:
\begin{eqnarray}
\label{before1} ds_{IIA}^2 =&  g_1~[dz  +
\Delta_1~\mbox{cot}~\hat{\theta}_1~(dx - b_{x\theta_1}~d\theta_1)
+ \Delta_2~\mbox{cot}~ \hat{\theta}_2~ (dy -
b_{y\theta_2}~d\theta_2)]^2 \\ \nonumber &  +  g_2~[d\theta_1^2 +
(dx - b_{x\theta_1} d\theta_1)^2] + g_3~ [d\theta_2^2 + (dy -
b_{y\theta_2} d\theta_2)^2{]}+ \\ \nonumber & +  g_4~{\rm sin}~\psi~ {[} (dx -
b_{x\theta_1}d\theta_1)d \theta_2 + (dy -
b_{y\theta_2} d\theta_2) d\theta_1 {]} + \\ \nonumber & + g_4~\mbox{cos}~\psi~ [
d\theta_1 d\theta_2 - (dx - b_{x\theta_1} d\theta_1) (dy -
b_{y\theta_2} d\theta_2)].
\end{eqnarray}
This metric (\ref{before1}) is exactly a non-K\"ahler deformation
of the metric for D6 branes on the three cycle of a deformed
conifold. The non-K\"ahler deformation can be seen from the
presence of the fields $b_{x\theta_1}$ in $d\hat{x} = dx - b_{x\theta_1}~d\theta_1$
and of $b_{y\theta_2}$ in $d \hat{y}=dy - b_{y\theta_2}~d\theta_2$. This
implies that the K\"ahler form $J$ and the 3-form $\Omega$ are
not-closed. Even though the manifold does not have an SU(3)
holonomy it has an SU(3) structure, so that supersymmetry is
preserved.

Furthermore one can easily show that any $B_{NS}$ field appearing
on the type IIA side after mirror is a gauge artifact. This is
most transparent if one chooses integrable complex structures for
the two tori in type IIB theory from the very beginning. There are
also non-trivial one form fluxes from the D6 branes sources. The
three form field vanishes and the coupling constant is equal to
the type IIB coupling constant.

This is the starting point of the type IIA transition. In order
to go to the closed string side, we need to first lift the
geometry to M theory to perform a flop and then dimensionally
reduce again to the type IIA theory \cite{amv}. The fact that the
type IIA theory is compactified on an SU(3)-structure manifold
implies that M theory is compactified on a $G_2$ structure
manifold. The absence of a $G_2$ holonomy is due to the
non-closure of $\Phi =J \wedge e^7 + \Omega_+ $ and its Hodge
dual. It was shown in \cite{trans1} that the identification of the
one forms and the performance of a flop can be done using the
methods of \cite{amv} \footnote{As expected, the one forms that we
would now require will be different from the ones chosen by
\cite{amv}. Indeed this is what we get. The one forms that we use
to specify the M-theory manifolds reduce to the one forms of
\cite{amv} when we turn off the non-K\"ahlerity of the type IIA
theory.}. After doing so and descending to the type IIA theory,
the result we get is
\begin{eqnarray}
ds^2 = h_1~ [d\theta_1^2 + (dx - b_{x\theta_1}~d\theta_1)^2]  +
h_2~[d\theta_2^2 + (dy - b_{y\theta_2}~d\theta_2)^2]  \\ \nonumber
+ h_3 ~[dz + \Delta_1~\mbox{cot}~\hat{\theta}_1~(dx -
b_{x\theta_1}~d\theta_1) + \Delta_2
~\mbox{cot}~\hat{\theta}_2~(dy - b_{y\theta_2}~d\theta_2)]^2,
\end{eqnarray}
 which is precisely the metric
of a resolved conifold when we switch off $b_{x\theta_1}$ and
$b_{y\theta_2}$. This is thus the closed string background with
no D6 branes but only sources. The manifold is non-K\"ahler as
well as non-complex. The identification between the open string
side and the closed string side was made by mapping the
expectation value of the gluino condensate on the stack of D6
branes and the volume of the resolution two cycle on the resolved
conifold side. This map requires the term  $J \wedge B_{(4)}$ on
the flux side \cite{vafa1}. Our proposal is that the presence of
$B_{(4)}$ is due to the fact that $d\Omega \ne 0$ and that the
type IIA superpotential contains a term $J \wedge d\Omega$
\footnote{For half-flat manifolds this has also been proposed in
\cite{wal}.}.

Making a further mirror transformation to this background we
obtain the closed string side of the type IIB geometric transition
\cite{trans2}. The type IIB manifold turns out to be a K\"ahler
deformed conifold with RR and NS three forms. This is exactly
what was expected from the results of \cite{vafa1}.

To summarize, non-K\"ahler manifolds play a crucial in the gauge theory/
string theory duality. They provide
important contributions to the superpotentials
which are crucial for a correct description of
the corresponding effective field theories.

\section{Type I/Heterotic Strings and  Non-Calabi-Yau Manifolds}

Even though non-K\"ahler manifolds were never studied in the
traditional string theory literature in much detail, their
importance has become evident in recent times due to the large
amount of new results in the area of string compactifications
with fluxes. We will now extend the calculation done in the
previous section to other type of models.

To do so, we start again from the type IIB compactification with
NS and RR fluxes and go to its orientifold limit which will
contain D7 branes and O7 planes. In order to obtain a metric, we
consider the metric of (\ref{2b}) and analyze which terms are
invariant under the O7 action. We consider the directions $x$ and
$y$ to be transverse to the O7 planes. The metric should then be
invariant under the orientifold action, should preserve some
number of supersymetries (i.e ${\cal N} = 1$), should have a form
close to the original type IIB metric and should allow wrapped D5
branes along with some number of D7 branes and O7 planes. Finally,
after two T-dualities, it's form should closely resemble the
metric obtained after T-dualizing the resolved conifold.

A metric which satisfies these conditions was computed in
\cite{trans2}
\begin{equation}
ds^2 = a_1(dx^2 + \vert \tilde\tau_1 \vert^2 ~dy^2 + 2
\mbox{Re}~\tilde\tau_1~dx~dy) + a_2 (d\theta_1^2 + \vert
\tilde\tau_2\vert^2~d\theta_2^2) + a_3~dz^2 + a_4~dr^2.
\end{equation}

After two T-dualities, this metric becomes a type I metric which takes the form
\begin{eqnarray}
ds^2 & = \alpha(1 + A^2)(dy - b_{y\theta_2}~d\theta_2)^2 +
\alpha (1 + B_1^2)(dx + b_{x\theta_1}~ d\theta_1)^2 +
\gamma'\sqrt{H} dr^2 \\ \nonumber ~& + 2~ \alpha A B ~(dx +
b_{x\theta_1}~d\theta_1)(dy - b_{y\theta_2}~d\theta_2) + \alpha
(1 + A^2) dz^2 + a_2 \vert d\chi \vert^2.
\end{eqnarray}
The T-dualities were performed along the $x$ and $y$ directions.
As $y$ is the angular direction of the $P^1$ cycle on which the D5
branes are wrapped on, the D5 branes loose the $y$ direction and
gain the direction $x$, thus the final configuration is again with
D5 branes wrapped on a $P^1$ cycle with the angular direction now
being the $x$ direction. Therefore, the starting point of the
geometric transition is given by D5 branes wrapping on a two cycle
inside a non-K\"ahler manifold.

After the transition we are again in the orientifold limit of some type IIB
configuration. If we impose the same conditions as we did before the
metric will take the form:
\begin{equation}
ds^2 = {b}_1~\vert d\chi_1 \vert^2 + {b}_2 ~\vert d\chi_2 \vert^2 + {b}_3 ~dz^2 +
 {b}_4~ dr^2,
\end{equation}
but now the complex structures will be different. We have
$\mbox{Re}~\tilde\tau_1 = 0$ and $\mbox{Re}~\tilde\tau_2 \ne 0$
for this solution while earlier the complex structures satisfied
$\mbox{Re}~\tilde\tau_1 \ne 0$ and $\mbox{Re}~\tilde\tau_2 = 0$.
The final type I manifold can be shown to be another non-K\"ahler
manifold \cite{trans2} whose explicit metric is given by:
\begin{eqnarray}
ds^2 & =  \frac{1}{h_2 + a_2^2 h_1} ~(dy - b_{y\theta_2} ~d\theta_2)^2 +
\frac{1}{h_4 + a_1^2 h_1} ~(dx - b_{x\theta_1}~ d\theta_1)^2 \\ \nonumber &
+ h_1~ dz^2 + h_3~h_4 ~\vert d\chi_2\vert^2 + \gamma'\sqrt{H}~ dr^2.
\end{eqnarray}
One important aspect of these type I compactifications is that
the metrics are all non-K\"ahler but complex. The integrability
of the complex structures is related to the torsional constraints
the metrics are required to satisfy, as well as the DUY equations
for the vector bundles.

We can go one step further by performing an S-duality to go from
the type I theory to the heterotic string. This means trading the
RR flux of the type I theory with the NS flux of heterotic. A
geometric transition will then take place between NS branes
wrapped on some two cycles of a non-K\"ahler complex manifold and
NS flux on another non-K\"ahler complex manifold.

One important check of our approach is the fact that there are two
conditions which should be mapped into one another.
These two conditions are the self-duality of the type IIB fluxes in the
orientifold limit
\begin{equation}
H_{RR} = * H_{NS},
\end{equation}
and the torsional equation for the heterotic string compactifications
\begin{equation}
H_{NS}= * dJ \equiv i~(\partial - \bar{\partial}) J.
\end{equation}
In \cite{trans2} it was shown that this mapping holds up to some
conjectured identification between constants which enter in the
flux definitions. This gives a strong check on the consistencies
of these results.

\section{Conclusions}

In this short note we summarized the complete duality cycle of the
geometric transitions taking place in all string theories and
M-theory. For the type II and M-theory transitions, these
transitions are well accounted for with various direct and
indirect checks. On the other hand the type I and heterotic cases
are relatively new. Our conjecture here is that the world volume dynamics
of wrapped branes in the type I and heterotic theories on
non-K\"ahler complex manifolds will have a description purely in terms of
another non-K\"ahler complex manifold with branes replaced by
fluxes. The manifolds predicted in the type I and heterotic
theories are examples of new non-K\"ahler complex manifolds that
complement the existing examples in the literature
\cite{prok,bbdgs,serone}.

\section*{Acknowledgments}

We thank S.~Alexander and A.~Knauf for their collaborations.
M.B. is supported by NSF
grant PHY-01-5-23911 and an Alfred Sloan Fellowship. K.B is
supported by NSF grant PHY-0244722, an Alfred Sloan Fellowship and
the University of Utah. K.D. is supported by
a Lucile and Packard Foundation Fellowship 2000-13856. R.T. is
supported by DOE Contract DE-AC03-76SF0098 and NSF grant
PHY-0098840. R.T. would like to thank KITP for hospitality during the
completion of this work. This research was supported in part by
National Science Foundation Grant No. PHY99-07949.

\end{document}